\documentclass[12pt,letterpaper]{article}

\usepackage[utf8]{inputenc}
\usepackage[english]{babel}
\usepackage{amsmath}
\usepackage{amsfonts}
\usepackage{amssymb}
\usepackage{graphicx}
\usepackage[left=1in,right=1in,top=1in,bottom=1in]{geometry}
\usepackage{changepage}
\usepackage{times}
\usepackage{cite}
\usepackage[colorlinks=true,citecolor=cyan,linkcolor=red,bookmarks=true,bookmarksnumbered=true]{hyperref}
\usepackage[font=footnotesize,labelfont=bf]{caption}
\usepackage{tikz}
\usetikzlibrary{arrows.meta}
\usepackage{subfigure}

\begin{document}
% Título
{
\flushleft\huge\bf High-order SUSY-QM, the Quantum XP Model and zeroes of the Riemann Zeta function 
}\vspace{5mm}

%% Información de los autores 
\begin{adjustwidth}{1in}{} %% Este entorno modifica el ancho del texto
{% Nombres
\flushleft\large\bf Juan D Garc{\'i}a-Mu{\~n}oz$^{1}$, A Raya$^{1,2}$ and Y Concha-S{\'a}nchez$^{3}$ 
}
\vspace{2.5mm}
{% Institución y dirreción
\par\noindent\small $^{1}$Instituto de F{\'i}sica y Matem{\'a}ticas, Universidad Michoacana de San Nicolás de Hidalgo, Edificio C-3, Ciudad Universitaria, Francisco J. M{\'u}jica S/N Col. Fel{\'i}citas del R{\'i}o, 58040 Morelia, Michoac{\'a}n, M{\'e}xico.
\par\noindent\small $^{2}$ Centro de Ciencias Exactas - Universidad del Bío-Bío. Avda. Andr\'es Bello 720, Casilla 447, Chillán, Chile.
\par\noindent\small $^{3}$Facultad de Ingeniería Civil, Universidad Michoacana de San Nicolás de Hidalgo, Edificio C, Ciudad Universitaria, Francisco J. Mújica S/N. Col. Felícitas del Río. 58030, Morelia, Michoacán, Mexico.
}
\vspace{2.5mm}
{% Correo electrónico
\par\noindent\footnotesize Email: juan.domingo.garcia@umich.mx, alfredo.raya@umich.mx and yajaira.concha@umich.mx 
}
\vspace{5mm}
{%% abstract
\par\noindent\bf Abstract
}
{ %% Escribir aquí el abstract
\newline\small     
Making use of the first- and second-order algorithms of supersymmetric quantum mechanics (SUSY-QM), we construct quantum mechanical Hamiltonians whose spectra are related to the zeroes of the Riemann Zeta function $\zeta(s)$. Inspired by the model of Das and Kalauni (DK) \cite{DAS2019265}, which corresponds to this function in the strip $0<Re[s]<1$, and taking the factorization energy equal to zero, we use the wave function $|x|^{-S}$, $S\in\mathbb{C}$, as a seed solution for our algorithms, obtaining XP-like operators. Thus, we construct SUSY-QM partner Hamiltonians whose zero energy mode locates exactly the nontrivial zeroes of $\zeta(s)$ along the critical line $Re[s]=1/2$ in the complex plane. We further find that unlike the DK case, where the SUSY-QM partner potentials correspond to free particles, our partner potentials belong to the family of inverse squared distance potentials with complex couplings.}
\vspace{2mm}
{%% keywords
}
\end{adjustwidth}
\vspace*{0.75cm}

\section{Introduction}\label{intro}

Riemann's hypothesis still remains as one of the challenging open questions in mathematics. The $\zeta(s)$ function defined as~\cite{riemann1860anzahl}
\begin{equation} \label{E-1}
\zeta(s)=\sum_{n=1}^\infty \frac{1}{s^n}, \quad s=\sigma+i\lambda, \quad \sigma,\lambda\in \mathbb{R}, \quad Re[s]>1.
\end{equation}
is closely connected to the number of prime numbers less than a given upper bound. The zeroes of this function, conjectured of the form $s_n=1/2\pm i \lambda_n$, are thus connected to the distribution of prime numbers. Many mathematical properties of the $\zeta(s)$ function are discussed in Ref.~\cite{riemann1860anzahl}. On the other hand, an extensive review of the applications of this function in many branches of physics have been discussed in Ref.~\cite{RevModPhys.83.307}. In particular, 
quantum mechanics offers a fertile ground to explore the location of the zeroes of the Riemann $\zeta(s)$ function. The Hilbert-Polya suggestion that these zeroes are related to the spectrum of an operator of the form ${\cal R}=\frac12 {\cal I}+i{\cal H}$, where ${\cal H}$ is self-adjoint and thus can be regarded as the Hamiltonian of a physical system, has driven a lot of attention towards a physical resolution of the Riemann hypothesis~\cite{doi:10.1137/S0036144598347497,berry1999h,Connes1999,doi:10.1142/S0219887817501092,doi:10.1142/S0219887818500950,PhysRevLett.118.130201,PhysRevA.101.043402,10.21468/SciPostPhysCore.4.4.032,Bender_2018,sierra2019riemann,Tamburini_2021,El-Nabulsi2019,https://doi.org/10.48550/arxiv.2002.12825,https://doi.org/10.48550/arxiv.2211.01899,bgsu1594035551136634}.

The famous XP-model developed by Berry, Keating and Connes~\cite{doi:10.1137/S0036144598347497,berry1999h,Connes1999} defined through  $H=xp$ provides an example of a Hamiltonian whose spectrum is closely related to the zeroes of the $\zeta(s)$ function. The regularization introduced by Berry and Keating~\cite{doi:10.1137/S0036144598347497,berry1999h} smoothly leads to a spectrum connected with the zeroes of $\zeta(s)$, whereas the variant introduced by Connes~\cite{Connes1999}, which is physically realizable as a system of charged particles restricted to move on a plane immersed in an electrostatic field  and a perpendicular magnetic field~\cite{PhysRevLett.101.110201}, leads to an absorption spectrum related to these zeroes as missing spectral lines. The XP-model has been revisited by several authors, as accounted for, for instance, in Ref.~\cite{sierra2019riemann}, where a modification of the Hamiltonian is introduced such that the spectrum is connected to the motion of massless Dirac particles in curved space-time.   

In a slightly different setting, Das and Kalauni~\cite{DAS2019265} constructed a first-order supersymmetric quantum mechanical (SUSY-QM) model, which we refer to as the DK model, that is intimately connected with $\zeta(s)$. By demanding that the the ground state energy of one of the supersymmetric partner potentials vanishes, the model naturally fixes the position of the zeroes of the Riemann Zeta function along the critical line. In this article we base our discussion along these findings, but extending the model through first- and second-order confluent supersymetric algorithms. Similar conclusions regarding the location of the zeroes are drawn from our models, but obtained from an entirely different algebra. To stress the similarities and differences of the DK model and the extensions proposed in this article, aiming for a self-contained discussion, the organization of the remaining of this manuscript is as follows: In Sect.~\ref{SUSY} we present the generalities of the SUSY-QM algorithm for a self-contained presentation of the problem. We present our model in Sect.~\ref{modelos} whereas a summary and outlook is left to the final Sect.~\ref{sum}.

\section{Supersymmetric Quantum Mechanics}\label{SUSY}
In Quantum Mechanics there exists an operational intertwining between two operators $H^{\pm}$, which is described by means of the rules
\begin{equation}\label{E-2}
    \left\{Q^{+},Q^{-}\right\}=H_{SS},\quad \left[Q^{\pm},H_{SS}\right]=0.
\end{equation}
It is standard to represent this algebra in terms of $2\times2$ matrices as follows
\begin{equation}\label{E-3}
    Q^{+}=
    \begin{pmatrix}
    0 & L^{+} \\
    0 & 0 
    \end{pmatrix},\quad 
    Q^{-}=
    \begin{pmatrix}
        0 & 0 \\
        L^{-} & 0
    \end{pmatrix},\quad
    H_{SS}=
    \begin{pmatrix}
        L^{+}L^{-} & 0 \\
        0 & L^{-}L^{+}
    \end{pmatrix} = 
    \begin{pmatrix}
        H^{-} & 0 \\
        0 & H^{+}
    \end{pmatrix},
\end{equation}
with $Q^{\pm}$ being the so-called superchargers, $H_{SS}$ the supersymmetric Hamiltonian and $L^{\pm}$ the intertwining operators\cite{Witten1981}. From the previous expressions, it is easy to arrive at the intertwining relation given by
\begin{equation}\label{E-4}
    H^{+}L^{-} = L^{-}H^{-}.
\end{equation}
If the Hamiltonians $H^{\pm}$ have eigenfunctions $\psi_{n}^{\pm}$ associated to eigenvalues $E^{\pm}_{n}$, from the intertwining relation in eq.~ \eqref{E-4}, we obtain that 
\begin{equation}\label{E-5}
    E_{n}^{+}=E_{n}^{-},\quad \psi_{n}^{\pm} = L^{\mp}\psi^{\mp}_{n},
\end{equation}
in other words, the eigenvalues of $H^{-}$ are also eigenvalues of $H^{+}$. When the supersymmetric algebra in eq.~\eqref{E-2} is satisfied, the Hamiltonians $H^{\pm}$ are called supersymmetric partners. It is important to note that this symmetry can take place for any two Hamiltonian operators as long as we can determine the corresponding intertwining operators $L^{\pm}$.

\subsection{First-order SUSY-QM}

Typically, SUSY-QM is introduced by taking two Schr{\"o}dinger-like Hamiltonians of the form\cite{Gangopadhyaya2018,Junker2019}
\begin{equation}\label{E-6}
    H_{1}^{\pm} = -\frac{d^{2}}{dx^{2}} + V_{1}^{\pm}(x),
\end{equation}
while the intertwining operators $L^{\pm}_{1}$ are first-order differential operators given by
\begin{equation}\label{E-7}
    L_{1}^{\pm} = \mp\frac{d}{dx} + w(x), 
\end{equation}
where the function $w(x)$ is referred to as the superpotential. By using the intertwining relation in eq.~\eqref{E-4}, it is possible to write the potentials $V^{\pm}_{1}$ in terms of the superpotential $w(x)$ as follows
\begin{equation}\label{E-8}
    V^{\pm}_{1} = w^{2}(x) \pm w'(x),
\end{equation}
with $f'(x)\equiv df(x)/dx$. Moreover, the superpotential is given by the seed function $u_{1}(x)$, which is a solution of the eigenvalue equation for $H^{-}$ associated to the factorization energy $\epsilon_{1}$, i.e.,
\begin{equation}\label{E-9}
    w(x) = -\frac{u'_{1}(x)}{u_{1}(x)},\qquad H^{-}u_{1}(x) = \epsilon_{1} u_{1}(x). 
\end{equation}
It is worth mentioning that the seed solution $u_{1}(x)$ must be a nodeless function inside the $x$-domain and the factorization energy $\epsilon_{1} \leq E_{0}$. Furthermore, the eigenfunction $\psi^{+}_{\epsilon_{1}}$ of the Hamiltonian $H^{+}$ associated to the factorization energy $\epsilon_{1}$ can be written as
\begin{equation}\label{E-10}
    \psi^{+}_{\epsilon_{1}} = \frac{1}{u_{1}(x)}.
\end{equation} 

\subsection{Confluent second-order SUSY-QM}
Confluent algorithm is a particular case of second-order Supersymmetric Quantum Mechanics, which is an algebraic method intertwining two Schr{\"o}dinger-like Hamiltonians $H_{2}^{\pm}$, whose form is analogous to the operators in eq.~\eqref{E-6}. In this case, an intertwining relation, similar to the one written in eq.~\eqref{E-4}, can be established. However, the intertwining operators $L^{\pm}_{2}$ are second-order differential operators. Specifically, they have the following form
\begin{equation} \label{E-11}
    L^{-}_{2} = \frac{d^{2}}{dx^{2}} + \eta(x)\frac{d}{dx} + \gamma(x),\quad L^{+}_{2} = (L^{-}_{2})^{\dagger},
\end{equation}
with $\eta(x)$ and $\gamma(x)$ being functions to be determined \cite{Nicolas2005} (see also \cite{Andrianov1993,Andrianov1995,Salinas2003,Salinas2005,Barnana2020,cs15a,cs15b,be16,cs17,sy18,bff12}). By substituting the expressions from eq.~\eqref{E-10} in the corresponding intertwining relation, in a straightforward way it is obtained that
\begin{equation} \label{E-12}
V^{+}_{2} = V^{-}_{2} + 2\eta',\quad \gamma = \frac{\eta^{2}}{2} - \frac{\eta'}{2} - V^{-}_{2} + \epsilon_{2},\quad V^{-}_{2} = \frac{\eta''}{2\eta} - \left(\frac{\eta'}{2\eta}\right)^{2} - \eta' + \frac{\eta^{2}}{4} + \epsilon_{2}.
\end{equation}
For simplicity, in the above equations we omitted the dependence of the functions on $x$. In the previous equations, we consider the constant $\epsilon_{2}$, which is the so-called factorization energy associated to the seed solution $u_{2}(x)$, fulfilling the stationary Schr{\"o}dinger-like equation for $H^{-}_{2}$, i.e., 
\begin{equation} \label{E-13}
    -u''_{2} + V^{-}_{2}u_{2} = \epsilon_{2} u_{2}.
\end{equation}
It is worth mentioning that the confluent algorithm is defined by means of the function $\eta$. In this case, that said function can be written as  
\begin{equation} \label{E-14}
    \eta = - \frac{\text{w}'}{\text{w}},\qquad \text{w} = \text{w}_{0} - \int\limits_{x_{0}}^{x}u^{2}_{2}(y)dy,
\end{equation}
where $x_{0}$ is a point in the appropriate $x$-domain and $\text{w}_{0}$ is a parameter that guarantees the function $\text{w}(x)$ remains nodeless. Furthermore, the eigenfunction $\psi^{+}_{\epsilon}(x)$ of the Hamiltonian $H^{+}_{2}$ corresponding to the factorization energy $\epsilon$ is directly proportional to 
\begin{equation} \label{E-15}
    \psi^{+}_{\epsilon} \propto \frac{u}{\text{w}}.
\end{equation}
Finally, note that confluent second-order SUSY-QM has an algebra defined by the same rules in eq.~\eqref{E-2}. However, the supersymmetric Hamiltonian $H_{SS}$ takes the following form 
\begin{equation} \label{E-16}
    H_{SS} = 
    \begin{pmatrix}
        L_{2}^{+}L_{2}^{-} & 0 \\
        0 & L_{2}^{-}L_{2}^{+}
    \end{pmatrix} = 
    \begin{pmatrix}
        (H^{-})^{2} & 0 \\
        0 & (H^{+})^{2}
    \end{pmatrix},
\end{equation}
while the supercharges $Q^{\pm}$ have the same form as in eq.~\eqref{E-3}, but now in terms of the operators $L_{2}^{\pm}$. Such an algebra is known as quadratic supersymmetric algebra. 
%In the next section, we construct a confluent SUSY-QM model for which the condition of the existence of zero modes leads naturally to the location of the zeroes of the Riemann $\zeta(s)$ function.
\section{Connection with the zeroes of the Zeta function} \label{modelos}
In this Section, we develop an algorithm to obtain pairs of SUSY partner Hamiltonians, no necessarily of the Schr{\"o}dinger-like form, whose eigenvalues are directly proportional to the Riemann Zeta function $\zeta(s)$. As a motivation, we first review the the DK model~\cite{DAS2019265} and then, we extend it by including an intertwining between two Schr{\"o}dinger-like Hamiltonians to determine operators with the function $\zeta(s)$ as eigenvalue.    

\subsection{The DK Model}
The DK model is based on the observation that monomials of the form $x^{-s}$ are eigenfunctions of the operators
\begin{equation} \label{E-17}
O_{DK}^{-} = \sum_{n=1}^\infty (-1)^{n+1}\exp{\left((\ln n)x \frac{d}{dx} \right)},\quad O^{+}_{DK} = \sum_{n=1}^\infty \frac{(-1)^{n+1}}{n}\exp{\left((\ln n^{-1})x \frac{d}{dx} \right)},
\end{equation}
namely,
\begin{equation} \label{E-18}
\begin{aligned}
O_{DK}^{-}x^{-s}&=(1-2^{1-s})\zeta(s)x^{-s}, \qquad Re[s]>0, \\
O^{+}_{DK} x^{-s}&=(1-2^{s})\zeta(1-s)x^{-s}, \qquad Re[s]<1.
\end{aligned}
\end{equation}
Thus, defining the raising and lowering operators
\begin{equation} \label{E-19}
    A^{-}(\omega)=|x|^{-i\frac{\omega}{2}}O_{DK}^{-} |x|^{-i\frac{\omega}{2}},\quad A^{+}(\omega)= |x|^{i\frac{\omega}{2}} O^{+}_{DK} |x|^{i\frac{\omega}{2}},
\end{equation}
the DK model is defined as the pair of Hermitian SUSY partner Hamiltonians
\begin{equation} \label{E-20}
H^+_{DK}=A^{-}(\omega)A^{+}(\omega),\quad H^-_{DK}=A^{+}(\omega)A^{-}(\omega),
\end{equation}
which together with the operators $A^{\pm}(\omega)$ as intertwining operators satisfy the supersymetric algebra in eq.~\eqref{E-2}. The ground state $\psi_0(x)$ of $H^-_{DK}$, defined such that $A^{-}(\omega)\psi_0(x) =0$, which is a zero mode. Explicitly,
\begin{equation} \label{E-21}
    \psi_0(x)=|x|^{-\frac{1}{2}+i(\frac{\omega}{2}-\lambda_*)},
\end{equation}
with the requirement of vanishing of the ground state energy fixes $\lambda_*=\omega/2-\rho$ to the position of a zero of $\zeta(s)$ along the critical line, i.e.,
\begin{equation} \label{E-22}
    \zeta\left(\frac12+i\lambda_*\right)=0.
\end{equation}

\subsection{Extensions of the DK model}\label{model}

\subsubsection{First-order SUSY}
Let us take the monomial $x^{-S_{0}}$, $S_{0} \in \mathbb{C}$, $x > 0$, as the seed solution $u(x)$ defining a first-order SUSY transformation with factorization energy $\epsilon = 0$. From eq.~\eqref{E-9}, we have that the superpotential is given by
\begin{equation} \label{E-23}
    w(x) = -\frac{u'(x)}{u(x)} = \frac{S_{0}}{x}.
\end{equation}
It is worth noting for $x < 0$ analogous results can be found, thus, we work in the range $(0,\infty)$ but the results can be written in terms of $|x|$ along the real line. In the traditional Schr{\"o}dinger form, the partner potentials are
\begin{equation} \label{E-24}
  V_{1}^{\pm}(x)=w^{2}(x) \pm w'(x) = \frac{S_{0}(S_{0}\mp 1)}{x^{2}}.
\end{equation}
Then, the Hamiltonians $H_{1}^{\pm}$ can be written as follows
\begin{equation} \label{E-25}
\begin{aligned}
    &H^{+}_{1} = L^{-}L^{+} = \left(\frac{d}{dx} + \frac{S_{0}}{x}\right)\left(-\frac{d}{dx}+\frac{S_{0}}{x}\right),\\
    &H^{-}_{1} = L^{+}L^{-} = \left(-\frac{d}{dx} + \frac{S_{0}}{x}\right)\left(\frac{d}{dx}+\frac{S_{0}}{x}\right).
\end{aligned}
\end{equation}
The way to connect with the Zeta function $\zeta(s)$ is to transform the previous Hamiltonians in operators factorized by the product $xp$. Thus, multiplying the Hamiltonians $H^{\pm}_{1}$ by $x^{2}$, we get that 
\begin{equation} \label{E-26}
\begin{aligned}
    &x^{2}H^{+}_{1} = \left(x\frac{d}{dx}+S_{0}-1\right)\left(-x\frac{d}{dx}+S_{0}\right),\\
    &x^{2}H^{-}_{1} = \left(-x\frac{d}{dx}+S_{0}+1\right)\left(x\frac{d}{dx}+S_{0}\right).
    \end{aligned}
\end{equation}
The monomials $x^{-S}$ are eigenfunctions of the operators written in the previous expressions, with eigenvalues $E^{\pm}_{S} = (S_{0}\pm S)(S_{0}\mp S\mp 1)$. The fact that these eigenvalues are different is an apparent contradiction with eq.~\eqref{E-5}. However, we must observe that from eq.~\eqref{E-10} the solution $\psi_{0}^{+} = x^{S_{0}}$ does not belong to the same monomial family. Thus, the operators in eq.~\eqref{E-26} have an isospectral part, in agrement with eq.~\eqref{E-5}, and their spectra only differ for the zero energy state of $x^{2}H^{-}$.

Now, recalling that the Zeta function $\zeta(s)$ has the following equivalent definitions 
\begin{equation} \label{E-27}
\begin{aligned}
    &\zeta(s) = \frac{1}{1-2^{1-s}}\sum\limits_{n=1}^{\infty}\frac{(-1)^{n+1}}{n^{s}},\quad \text{Re} [s] > 0, \\ 
    &\zeta(1-s) = \frac{1}{1-2^{s}}\sum\limits_{n=1}^{\infty}\frac{(-1)^{n+1}}{n^{1-s}},\quad \text{Re} [s] < 1,
    \end{aligned}
\end{equation}
if we select, for example,  the factors whose product is equal to $x^{2}H^{-}_{1}$ in eq.~\eqref{E-26} (analogous results can be obtained for the factors whose product equals $x^{2}H^{+}_{1}$), it is possible to construct the following operators
\begin{equation}\label{E-28}
    \begin{aligned}
    &O_{1}^{-} = \sum\limits_{n=1}^{\infty}(-1)^{n+1}\exp\left[(\ln n)\left(x\frac{d}{dx}+S_{0}\right)\right],\\
    &O^{+}_{1} = \sum\limits_{n=1}^{\infty}\frac{(-1)^{n+1}}{n^{2}}\exp\left[(\ln n)\left(-x\frac{d}{dx}+S_{0}+1\right)\right],
    \end{aligned}
\end{equation}
which have as eigenfunctions the monomials $x^{-S}$, i.e.,
\begin{equation} \label{E-29}
\begin{aligned}
    &O_{1}^{-}x^{-S} = \left(1 - 2^{1 - S + S_{0}}\right)\zeta (S - S_{0})x^{-S},\quad \text{Re}[S - S_{0}] > 0, \\
    &O^{+}_{1}x^{-S} = \left(1 - 2^{S + S_{0}}\right)\zeta (1- S - S_{0})x^{-S},\quad \text{Re}[S + S_{0}] < 1.
\end{aligned}
\end{equation}
In order to link to the critical line of the Riemann Zeta function, we take $S = \sigma - i\rho$ and $S_{0} = i\rho_{0}$, and thus it turns out that
\begin{equation} \label{E-30}
    \begin{aligned}
        &O_{1}^{-}x^{-\sigma + i\rho} = \left(1 - 2^{1 - \sigma + i(\rho + \rho_{0})}\right)\zeta (\sigma - i(\rho + \rho_{0}))x^{-\sigma + i\rho},\quad \sigma > 0, \\
        &O_{1}^{+}x^{-\sigma + i\rho} = \left(1 - 2^{\sigma - i(\rho - \rho_{0})}\right)\zeta (1 - \sigma + i(\rho - \rho_{0}))x^{-\sigma + i\rho},\quad \sigma < 1.
    \end{aligned}
\end{equation}
Thus, we can construct Hamiltonians $H_{OM}^{\pm}$ in analogy to the DK model in eq.~\eqref{E-20}. Such Hamiltonians are given by
\begin{equation} \label{E-31}
    H_{OM}^{+} = A^{-}_{OM}(\omega)A^{+}_{OM}(\omega),\quad H_{OM}^{-} = A^{+}_{OM}(\omega)A^{-}_{OM}(\omega),
\end{equation}
where the operators $A^{\pm}_{OM}(\omega)$ are defined as follows
\begin{equation}\label{E-32}
    A^{\pm}_{OM}(\omega) = |x|^{\pm i\frac{\omega}{2}}O_{1}^{\pm}|x|^{\pm i\frac{\omega}{2}}.
\end{equation}
Since in the real positive line the functions $x^{-1/2+i\rho}$ are the unique kind of functions $x^{-S}$ being normalizable \cite{DAS2019265}, then, taking $\sigma = 1/2$, the actions of the Hamiltonians $H^{\pm}_{OM}$ into the functions $|x|^{-1/2 +i\rho}$ are
\begin{equation} \label{E-33}
    \begin{aligned}
        H^{\pm}_{OM}|x|^{-\frac{1}{2} + i\rho} &= \left(1-2^{\frac{1}{2}-i(\rho-\rho_{0}\pm\frac{\omega}{2})}\right)\left(1-2^{\frac{1}{2}+i(\rho+\rho_{0}\pm\frac{\omega}{2})}\right)\times\\
        &\times\zeta \left(\frac{1}{2}+i\left(\rho-\rho_{0}\pm\frac{\omega}{2}\right)\right)\zeta \left(\frac{1}{2}-i\left(\rho+\rho_{0}\pm\frac{\omega}{2}\right)\right)|x|^{-\frac{1}{2}+i\rho}.
    \end{aligned}
\end{equation}
Both Hamiltonians have a corresponding zero energy state, and it is straightforward to get that 
\begin{equation} \label{E-34}
    \zeta \left(\frac{1}{2}+i\left(\rho-\rho_{0}\pm\frac{\omega}{2}\right)\right)\zeta \left(\frac{1}{2}-i\left(\rho+\rho_{0}\pm\frac{\omega}{2}\right)\right) = 0.
\end{equation}
Hence, one of these Zeta functions vanishes, i.e., $\zeta (1/2 +i\lambda)=0$, where $\lambda$ is the localization of a zero on the critical line. 

It is worth mentioning the Hamiltonians $H_{OM}^{\pm}$ in eq.~\eqref{E-31} are non-Hermitian. However, our model can be reduce to the DK model by choosing $\rho_{0} = 0$, whereby $H_{OM}^{\pm} = H_{DK}^{\pm}$, which are Hermitian.  
% where the (complex) couplings are, respectively,
% \begin{equation} \label{E11}
%    \alpha_{-}= -\left(\rho+\frac{i}{2}\right)\left(\rho+\frac{3i}{2}\right),\quad \alpha_{+}=-\left(\frac{1}{4}+\rho^2 \right).
% \end{equation}
% We use these results to set up our model through the confluent algorithm of SUSY-QM. 

\subsubsection{Confluent second-order SUSY}
Making use of the confluent algorithm, we start by taking the seed solution as
\begin{equation} \label{E-35}
u = x^{-S_{0}},\qquad S_{0} \in \mathbb{C}.
\end{equation}
Once again, even though we take $x > 0$, analogous results can be obtained for $x < 0$, and the results can be written in terms of $|x|$ in the real line. Considering the integral
\begin{equation} \label{E-36}
    I = \int\limits_{x_{0}}^{x}u^{2}(y)dy = \frac{x^{1-2S_{0}} - x_{0}^{1-2S_{0}}}{1-2S_{0}},
\end{equation}
we have that
\begin{align} \label{E-37}
\text{w} = \text{w}_{0}-I = \text{w}_{0} + \frac{x_{0}^{1-2S_{0}}}{1-2S_{0}} - \frac{x^{1-2S_{0}}}{1-2S_{0}}
\end{align}
Since $x_{0}$ is a fixed point in the $x$-domain, we can choose $\text{w}_{0} = -(x_{0}^{1-2S_{0}})/(1-2S_{0})$. Thus, the function $\text{w}$ reduces to   
%In order to avoid zeroes in this function, we can see its imaginary part vanishes provided $x=x_{0}$. Nevertheless, there exist an infinite number of points where such an imaginary part would vanish, given the periodic nature of the functions involved. For simplicity, we consider the interval
\begin{equation} \label{E-38}
\text{w} = \frac{x^{1-2S_{0}}}{2S_{0}-1}.   
\end{equation}
Note that the function $\text{w}$ is nodeless inside the $x$-domain $(0,\infty)$. Then, the confluent transformation is suitable, and substituting the function $\text{w}$ in eq.~\eqref{E-14}, it is obtained that 
\begin{equation} \label{E-39}
\eta = \frac{2S_{0}-1}{x}.    
\end{equation}
% It becomes convenient to select the left corner of the interval such that $x_{0} = -e^{\frac{\pi}{4\rho}}$. Thus,
% \begin{equation} \label{E24}
%   \text{w}(x) = \text{w}_{0} + \frac{1}{2\rho} + i\frac{|x|^{i2\rho}}{2\rho}.  
% \end{equation}
Therefore, using the eq.~\eqref{E-12}, the potentials that come from the confluent transformation are
\begin{equation} \label{E-40}
      V^{-}_{2}(x) = \frac{S_{0}(S_{0}+1)}{x^{2}},  
  \qquad
    V^{+}_{2}(x) = \frac{(S_{0}-2)(S_{0}-1)}{x^2}.
\end{equation}
In this case, the intertwining operators have the form 
\begin{equation} \label{E-41}
\begin{aligned}
    &L_{2}^{-} = \frac{d^{2}}{dx^{2}} + \frac{2S_{0}-1}{x}\frac{d}{dx} + \frac{S_{0}(S_{0}-2)}{x^{2}}, \\
    &L_{2}^{+} = \frac{d^{2}}{dx^{2}} - \frac{2S_{0}-1}{x}\frac{d}{dx} + \frac{S_{0}^{2}-1}{x^{2}}.
\end{aligned}
\end{equation}
In order to link to the Riemann Zeta function $\zeta (s)$, we multiply the intertwining operators by $x^{2}$ on the right-hand side and therefore, 
\begin{equation}\label{E-42}
    \begin{aligned}
        &x^{2}L_{2}^{-} = x^{2}\frac{d^{2}}{dx^{2}} + (2S_{0}-1)x\frac{d}{dx} + S_{0}(S_{0}-2), \\
        &x^{2}L_{2}^{+} = x^{2}\frac{d^{2}}{dx^{2}} - (2S_{0}-1)x\frac{d}{dx} + S_{0}^{2}-1.
    \end{aligned}
\end{equation}
The previous operator can be factorized as follows
\begin{equation}\label{E-43}
    \begin{aligned}
        &x^{2}L_{2}^{-} = \left(x\frac{d}{dx}+S_{0}\right)\left(x\frac{d}{dx}+S_{0}-2\right),\\
        &x^{2}L_{2}^{+} = \left(x\frac{d}{dx}-S_{0}-1\right)\left(x\frac{d}{dx}-S_{0}+1\right).
    \end{aligned}
\end{equation}
Let us observe that the factors are XP-like operators. Hence, we can construct new operators $O_{OM2}^{\pm}$ whose eigenvalues are directly proportional to the Riemann Zeta function. For example, taking the corresponding factors of $x^{2}L_{2}^{+}$, we define the following operators
\begin{equation}\label{E-44}
\begin{aligned}
    &O_{OM2}^{-} = \sum\limits_{n=1}^{\infty}\frac{(-1)^{n+1}}{n}\exp\left[(\ln n)\left(x\frac{d}{dx} - S_{0} + 1\right)\right], \\
    &O_{OM2}^{+} = \sum\limits_{n=1}^{\infty}\frac{(-1)^{n+1}}{n^{2}}\exp\left[(\ln n^{-1})\left(x\frac{d}{dx} - S_{0} -1\right)\right],
\end{aligned}    
\end{equation}
whose actions into the monomials $x^{-\sigma+i\rho}$, with $S_{0} = i\rho_{0}$ are 
\begin{equation}\label{E-45}
    \begin{aligned}
        &O^{-}_{OM2}x^{-\sigma+i\rho} = \left(1-2^{1-\sigma+i(\rho-\rho_{0})}\right)\zeta \left(\sigma - i(\rho-\rho_{0})\right),\quad \sigma > 0,\\
        &O^{+}_{OM2}x^{-\sigma+i\rho} = \left(1-2^{\sigma-i(\rho-\rho_{0})}\right)\zeta \left(1-\sigma+i(\rho-\rho_{0})\right),\quad \sigma < 1.
    \end{aligned}
\end{equation}
\begin{figure}
    \centering
    \subfigure[]{\includegraphics[scale=0.5]{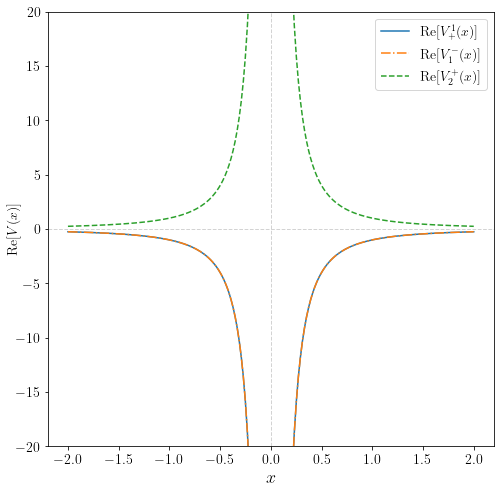}}
     \subfigure[]{\includegraphics[scale=0.5]{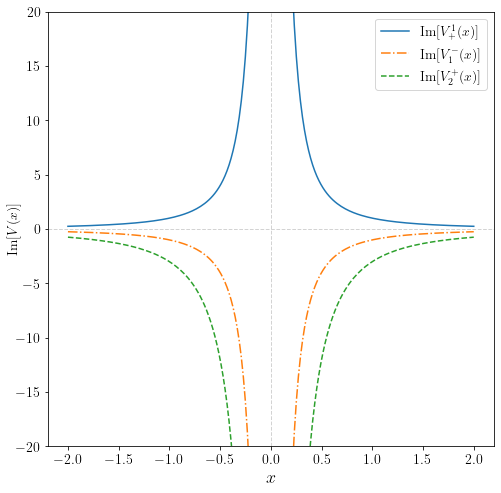}}
    \caption{(a) Real parts of the first-order supersymmetric partner potentials $V^{1}_{\pm}(x)$ and of the confluent supersymmetric partner potential $V^{+}_{2}(x)$. (b) Imaginary parts of the same potentials $V^{\pm}_{1}(x)$ and $V_{2}^{+}(x)$. Note that $V^{-}_{1}(x) = V^{-}_{2}(x)$. The scale of the graphs is set by the parameter $S_{0} = i\rho_{0}$, with $\rho_{0} = 1$.}
    \label{F1}
\end{figure}
Using a similar procedure to obtain the Hamiltonians $H_{OM}^{\pm}$ in eq.~\eqref{E-31}, we can get the Hamiltonians $H_{OM2}^{\pm}$ whose actions into the functions $|x|^{-1/2+i\rho}$ are given by
\begin{equation}\label{E-46}
    H_{OM2}^{\pm}|x|^{-\frac{1}{2}+i\rho} = \left|1-2^{\frac{1}{2}+i(\rho-\rho_{0}\pm\frac{\omega}{2})}\right|^{2}\left|\zeta\left(\frac{1}{2}+i\left(\rho-\rho_{0}\pm\frac{\omega}{2}\right)\right)\right|^{2}|x|^{-\frac{1}{2}+i\rho}.
\end{equation}
It is worth noting that these Hamiltonians are Hermitian. On the other hand, they have a corresponding zero energy state, and thus, such state is related to a zero of the Riemann Zeta function on the critical line, indeed,
\begin{equation}\label{E-47}
    \left|\zeta\left(\frac{1}{2}+i\left(\rho-\rho_{0}\pm\frac{\omega}{2}\right)\right)\right|^{2}=0 \qquad\Rightarrow\qquad \zeta \left(\frac{1}{2}+i\lambda\right) = 0, 
\end{equation}
where $\lambda = \rho-\rho_{0}\pm \omega/2$ indicates the position of the zero of the function $\zeta(s)$ on the critical line. Finally, it is interesting that despite the eigenvalues of the Hamiltonians $H_{OM2}^{\pm}$ in eq.~\eqref{E-46} reduce to the same values of the DK model when $\rho_{0} = 0$, the operators $O_{OM2}^{\pm} \neq O_{DK}^{\pm}$; while the confluent transformation defined by the function $\eta$ in eq.~\eqref{E-39} intertwines a free particle Hamiltonian $H^{-}_{2}$ with a Hamiltonian $H_{2}^{+}$ describing a quantum particle in a centrifugal potential.  
%Notice that~\eqref{E19} is a zero mode eigenstate of $V^-(x)$, and thus the vanishing of its energy eigenvalue also implies that the location of a zero of the Riemann zeta function is found as in Eq.~\eqref{E7}. Furthermore, we must mention that being rigorous, in Eqs.~\eqref{E10} and~\eqref{E25} we should write $|x|^{2}$ instead of $x^{2}$. However, this simplification is possible since $x\in\mathbb{R}$.

\section{Summary and outlook}\label{sum}
In summary, we have obtained both, first- and second-order SUSY-QM models which lead to XP-like operators. By using such operators it is possible to construct the Hamiltonians in eqs.~(\ref{E-33}, \ref{E-46}) having a zero mode fixing the location of the zeroes of the Riemann Zeta function $\zeta(s)$ along the critical line. At first sight, it might suggest that the models at hand are similar to those of the DK model, since our models also achieve SUSY partner Hamiltonians whose spectra are formed by eigenvalues directly proportional to the Zeta function. Nevertheless, one should stress that the first-order formalism discussed in this article leads to non-Hermitian Hamiltonians, whereas the second-order formalism, being in fact more robust, leads to Hermitian Hamiltonians that are in fact, much different from the DK Hamiltonians, thus showing that the family of Hamiltonians whose spectra connect with the Riemann Zeta function is extensive. Finally, the SUSY partner potentials $V^{\pm}_{1}(x)$ and $V^{\pm}_{2}(x)$ in eqs.~(\ref{E-24}, \ref{E-40}) (see also Fig.~\ref{F1}) exhibit a more intricate behavior as compared with the corresponding to the DK model, which correspond to free particle potentials. Many interesting features of the properties of bound states in $1/x^2$ potentials have been discussed, for instance, in Refs.~\cite{Essin2006,Nguyen2020} in several cases with real couplings. Extensions of these ideas for complex couplings as the ones derived in this article are under consideration and results would be presented elsewhere. 
 
\section*{Acknowledgenments}
We benefited from valuable discussions with Prof. Ashok Das. 
JDGM and AR acknowledge financial support from CONACYT Project FORDECYT-PRO{\-N}ACES/61533/2020. YCS also acknowledges the CIC-UMSNH research grant 6976882/2023.

\bibliographystyle{ieeetr}
\bibliography{biblio}
\end{document}